\documentclass[aps,prb,twocolumn,superscriptaddress]{revtex4}
\usepackage{amsmath}
\usepackage{amssymb}
\usepackage{graphicx}
\usepackage{bm}

\begin{document}

\preprint{LA-UR-08-2381}

\title{Two inequivalent sublattices and orbital ordering in MnV$_2$O$_4$ by 
$^{51}$V NMR}


\author{S.-H. Baek}
\affiliation{Los Alamos National Laboratory, Los Alamos, NM 87545}
\author{N. J. Curro}
\affiliation{Department of Physics, University of California, Davis, CA 95616}
\author{K.-Y. Choi}
\affiliation{Department of Physics, Chung-Ang University, Seoul 156-756, Republic of Korea} 
\author{A. P.  Reyes}
\affiliation{National High Magnetic Field Laboratory, Tallahassee, Florida
32310}
\author{P. L.  Kuhns}
\affiliation{National High Magnetic Field Laboratory, Tallahassee, Florida
32310}
\author{H. D.  Zhou}
\affiliation{National High Magnetic Field Laboratory, Tallahassee, Florida
32310}
\author{C. R.  Wiebe}
\affiliation{National High Magnetic Field Laboratory, Tallahassee, Florida
32310}

\date{\today}

\begin{abstract}
We report detailed $^{51}$V NMR spectra in a single crystal of MnV$_2$O$_4$.
The vanadium spectrum reveals two peaks in the orbitally ordered state, which 
arise from different internal 
hyperfine fields at two different V sublattices. 
These internal fields evolve smoothly with externally applied field, and show
no change in structure that would suggest a change of the orbital ordering.
The result is consistent with the orbital ordering model recently proposed by 
Sarkar \textit{et al.}~[Phys. Rev. Lett. \textbf{102}, 216405 (2009)] in 
which the same orbital that is a mixture of $t_{2g}$ orbitals rotates by about 
45$^\circ$ alternately within and between orbital  
chains in the $I4_1/a$ tetragonal space group. 
\end{abstract}

\pacs{}

\maketitle

Competition between spin interactions and orbital degeneracy is a key factor
in determining the ground state magnetic and lattice structure of transition
metal oxides, and gives rise to a rich spectrum of phase transitions in
magnetic insulators.\cite{kugel82,radaelli05}   Recently, vanadium
oxide spinels of the form AV$_2$O$_4$ have attracted interest, in which A
is a divalent transition element that is either non-magnetic\cite{mamiya97, 
ueda97, onoda03} (Mg, Zn, Cd) or
magnetic\cite{plumier89, adachi05, zhou07, suzuki07, garlea08} (Mn).
The intriguing physics of the vanadate compounds arises from the geometrical
spin frustration of the triply degenerate $t_{2g}$ orbitals of V$^{3+}$ ($3d^2$,
$S=1$) in the spinel structure. The V ions sit at the vertices of corner-sharing
tetrahedra, and experience a magnetic exchange interaction with nearest
neighbor V spins. In the process of relieving spin frustration, the
vanadates typically undergo two consecutive phase
transitions. First, a 
structural distortion splits the $t_{2g}$ levels into a
low-lying $xy$ orbital and a higher doublet ($yz$, $zx$) at a temperature $T_S$.
This structural phase transition is accompanied by
long-range orbital ordering of the V orbitals, since Hund's rules imply that
a single electron occupies the excited doublet and orbital exchange
interactions lift this degeneracy.  If the A site is non-magnetic, then the V
spins order antiferromagnetically at a lower temperature $T_N < T_S$.

\begin{figure}
\centering
\includegraphics[width=0.9\linewidth]{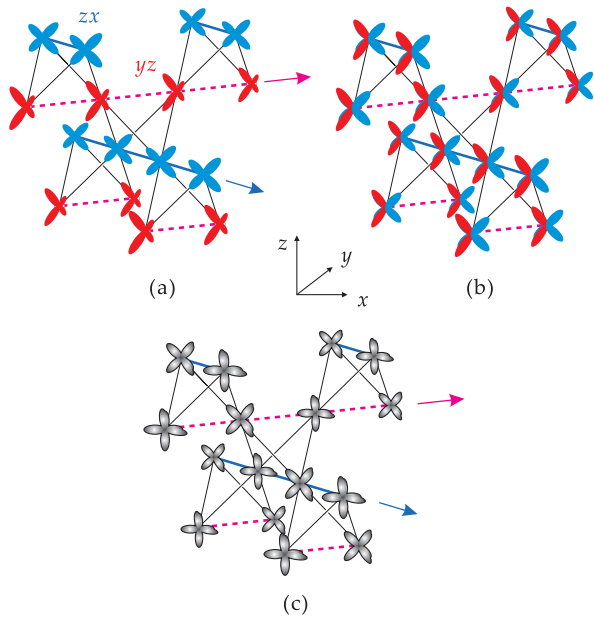}
\caption{(Color online) Proposed orbital ordering models: (a) staggered 
antiferro-orbital order in  
$I4_1/a$ symmetry,\cite{suzuki07, garlea08}
(b) ferro-orbital order in $I4_1/amd$ symmetry,\cite{adachi05} (c) 
``antiferro''-orbital order in $I4_1/a$ symmetry.\cite{sarkar09} Blue (solid) 
and red (dotted) lines represent orbital chains running along the edges, 
corresponding to [110],  of the corner-shared V tetrahedra.  
\label{fig:oo}}
\end{figure}

When the A site is magnetic, novel features emerge that are distinct from the
other vanadates.   In particular, for Mn$^{2+}$ ($3d^5$) a ferrimagnetic (FEM)
transition occurs \textit{before} the structural transition ($T_S < T_N$). The
combination of strong ferromagnetic Mn-Mn  couplings, antiferromagnetic (AFM) Mn-V
couplings, and AFM V-V couplings leads to a collinear
FEM spin configuration at $T_N \sim 56$ K. This collinear state
retains the orbital degeneracy, but is unstable so that upon further cooling
the V spins become non-collinear below the structural transition at $T_S = 53$
K. Once again, long range orbital order of the V $d$ orbitals emerges in concert
with structural distortion. However, the nature of the orbital symmetry 
remains poorly understood and 
continues to be debated theoretically in the literature.\cite{tsunetsugu03, tchernyshyov04,
matteo05, sarkar09} 
Experimentally, Adachi \textit{et al.}~proposed the ferro-orbital ordering in 
the $I4_1/amd$ tetragonal symmetry,\cite{adachi05} but 
later x-ray and neutron experiments on single crystals\cite{suzuki07, garlea08}  
claimed that the $I4_1/a$ space group was present with antiferro-orbital 
ordering in which $yz$ and $zx$ orbitals  
alternate along the $c$ axis. Although $I4_1/a$ space group allows only 
antiferro-orbital order due to the symmetry consideration, Chung et 
al.~\cite{chung08} pointed out that  
the large exchange coupling along $c$ determined by neutron scattering is 
contradictory with the simple antiferro-orbital ordering. In fact, taking into 
account the staggered trigonal distortion, recent  
first principles calculations proposed a new  
orbital ordering in which the same orbital forms ``antiferro'' orbital order 
by rotating its direction by $\sim 45^\circ$ alternately within and between 
orbital chains, while maintaining $I4_1/a$ symmetry.\cite{sarkar09} 
The newly proposed orbital ordering model reconciles the 
contradicting experimental results and, indeed, is in good agreement with our 
NMR results. The three proposed orbital ordering models are schematically 
drawn in Fig.~\ref{fig:oo}.

In order to investigate the orbital
and magnetic order microscopically, we have carried out $^{51}$V nuclear
magnetic resonance (NMR) as a function of field ($H$) and temperature ($T$) within the FEM
state.  In zero field, the $^{51}$V nuclear spins levels are split by an
internal hyperfine
field from the ordered Mn and V moments.  The field
dependence of the resonance reveals the non-collinear nature of the ordered
moments.  Single ferrimagnetic domain is formed above $H_c\sim0.3$ T, while 
the orbital ordering is intact up to 13 T.  

\begin{figure}
\centering
\includegraphics[width=0.9\linewidth]{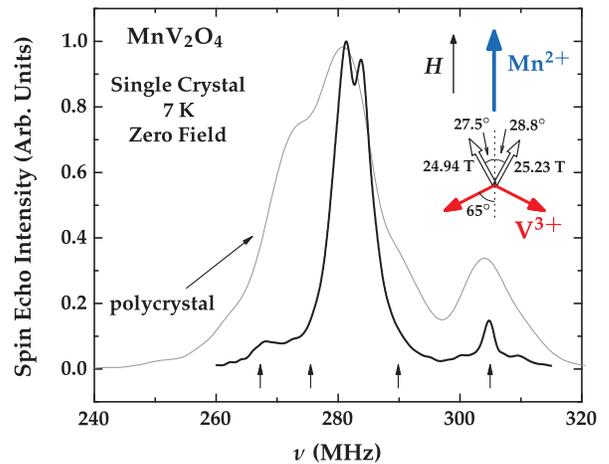}
\caption{(Color online) $^{51}$V spectrum of single crystal at zero field. Compared with
spectrum of polycrystal, the structure is much simple and narrow, indicating
that single crystal is relatively free of the impurity cubic phase.
The contribution of small impurity phases remaining in the 
single crystal are indicated by arrows. 
Two main peaks in single crystal spectrum implies the existence of inequivalent
V sublattices, as depicted in inset, in which hyperfine fields at V nuclei 
(empty arrows) are not antiparallel to the V$^{3+}$ spin moments. The heights 
of two spectra were nomalized for comparison.  
 \label{fig:spec} } 
\end{figure}

$^{51}$V NMR spectra were obtained on a single crystal of MnV$_2$O$_4$ between
4 and 35 K in zero field and in
external fields up to 13 T. The preparation of the single crystal of MnV$_2$O$_4$ 
has been described in detail in Ref.~\onlinecite{zhou07}. The spectra were
obtained by integrating averaged spin echo signals as the frequency
was swept through the resonance line. The V resonance in the single crystal (SC) is close to
285 MHz as we found in previous measurement of a polycrystal (PC) sample.\cite{baek08} 
However, the
spectrum in the SC is narrower than in the PC and reveals two sharp features,
whereas the PC sample is significantly broader with poorly resolved features
(Fig.~\ref{fig:spec}). These differences are consistent with recent x-ray
diffraction and magnetization measurements,\cite{suzuki07} as well as
specific heat data,\cite{zhou07} which show evidence of an impurity cubic phase resulting from
non-stoichiometric crystallites in
PC samples.\cite{zhou07} The spectrum of SC clearly shows
two narrow lines, but also reveals weak signals, in particular, near 270 MHz and
305 MHz, due to a small portion of the impurity cubic 
phase remained in SC sample. 

In the absence of quadrupolar effects, the V resonance frequency is given by
$\nu_0=\gamma_N H_\text{hf}$ where
$\gamma_N = 11.195$ MHz/T is the nuclear gyromagnetic ratio, and $H_\text{hf}$
is the hyperfine field at the V. \textit{A priori}, one
might expect that the two well resolved peaks might arise from the
second order quadrupolar contribution in the 
orbitally ordered state. In this case, the splitting can be written as\cite{bennet} 
$\Delta\nu=25\nu_Q^2/144\nu_0[I(I+1)-3/4]$ and the NQR frequency $\nu_Q$  
is estimated to be $\sim 40$ MHz. Since, however, we are unable to find any satellite 
transitions associated with the first order quadrupole splitting, we conclude that $\nu_Q$ 
is negligibly small and the 
origin of the two peaks is not quadrupolar in nature. Thus the two peaks must 
arise from two different 
\textit{hyperfine} fields at different V sublattices which exist only in a
state with both long-range orbital and spin order.

It is clear that there is only one V site in the cubic spinel structure above $T_S$.  
The proposed $I4_1/a$ tetragonal 
structure below $T_S$ exhibits a slight contraction of the VO$_6$ octahedra along the $c$
direction, but still retains a single crystallographic V site.\cite{garlea08}
However, the hyperfine fields at V nuclei can be differentiated. The net 
hyperfine field at the V can be  
written as the sum of on-site and transferred terms : 
$\mathbf{H}_\text{hf} = \mathbf{H}^\text{on-site}_\text{hf} + 
\mathbf{H}^\text{trans}_\text{hf}$. Each term can be decomposed as:

\begin{align}
\label{eq:hf0}
\mathbf{H}^\text{on-site}_\text{hf} & = \mathbf{H}_F + \mathbf{H}_l +  
\mathbf{H}_d,  \\ 
\mathbf{H}^\text{trans}_\text{hf} & = 
\sum_i A_i \mathbf{S}^i + \sum_j B_j \mathbf{S}_\text{Mn}^j,
\end{align}
where $\mathbf{H}_F$ is the Fermi contact 
field arising from the core polarization, $\mathbf{H}_l$ the orbital 
term from the orbital momentum $\mathbf{L}$, $\mathbf{H}_d$ the dipolar field 
from the on-site electrons,
the indices $i$ and $j$ are over nearest neighbors V and Mn spin moments, respectively.  
For an orbital triplet ion, the orbital momentum is only 
partially quenched\cite{abragam70} and could be as large as 0.34 
$\mu_B$ in MnV$_2$O$_4$,\cite{sarkar09} causing large $\mathbf{H}_l$ and $\mathbf{H}_d$. 
On the other hand, the transferred terms are usually very small compared  
to the on-site term,\cite{kikuchi96} but could be large enough to produce two peaks.

\begin{figure}
\centering
\includegraphics[width=0.9\linewidth]{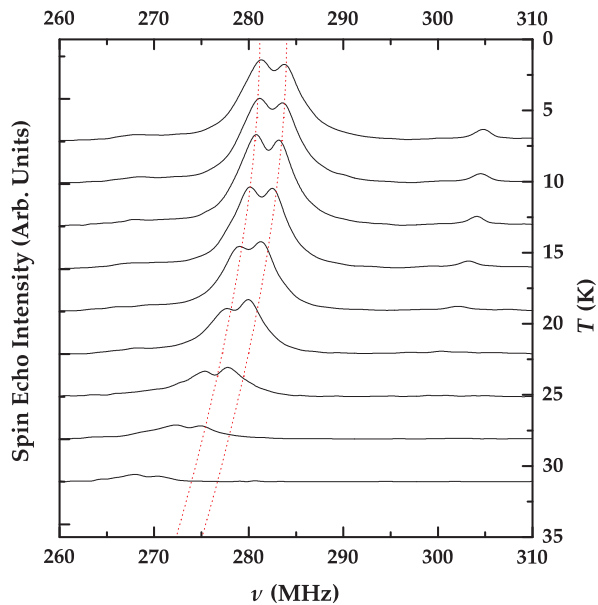}
\caption{(Color online) Temperature dependence of $^{51}$V spectrum in zero field.
A Boltzmann correction by multiplying $T$ was made. The signal
becomes weak rapidly with increasing $T$ and disappears above 35 K due to
the shortening of $T_2$ associated with thermal fluctuation. Dotted lines are
Eq.~(\ref{eq:sw}), implying that the spin wave theory is applicable at low $T$
region. Seemingly different $T$ dependence of relative intensities between two peaks is
ascribed to the impurity cubic phase in which resonance frequencies are almost
$T$ independent.\cite{baek08} \label{fig:Tspec}}
\end{figure}

In the case of the perfect antiferro-orbital 
order\cite{tsunetsugu03,suzuki07,garlea08} or ferro-orbital   
order,\cite{tchernyshyov04, adachi05} the hyperfine fields should be identical for all of 
the V sites. 
Therefore, our data suggest that (i) the tetragonal symmetry is lower than 
$I4_1/a$ to allow the two V sites, or (ii) the orbital ordering is somewhat
complex, yet without lowering the tetragonal symmetry $I4_1/a$. 
Case (i) is ruled out basing on the neutron and x-ray results\cite{garlea08, 
suzuki07} so that we assume that case (ii) is applicable. 
Indeed, case (ii) is realized in the orbital ordering model proposed by Sarkar 
et al.\cite{sarkar09} in which the same orbital rotates alternately within 
and between orbital chains by about 45$^\circ$.  
Since the model was obtained by taking into consideration the trigonal distortion of 
VO$_6$ octahedra in addition to the tetragonal contraction, it naturally 
explains the two V sublattices still in the $I4_1/a$  
tetragonal symmetry because the V-V  
and Mn-V transferred hyperfine fields become different for two V  
sublattices depending on the orbital overlap between V-V and/or V-Mn via oxygen 
$2p$ orbitals [see Fig.~\ref{fig:oo}(c)]. 

The two differentiated V sites appear to be robust against the variation of 
temperature and external field.   
As shown in Fig.~\ref{fig:Tspec},
with increasing $T$, the spectrum shifts to lower frequencies, while retaining 
the same shape.  Since 
$H_\text{hf}$ is proportional to the sublattice magnetization, we expect
$\nu(T)$ to follow Bloch's
$T^{3/2}$ law at $T\ll T_N$. If there is an energy gap, $E_g$, in the spin wave
excitation spectrum, $\nu(T)$ at sufficiently low $T$ is given by:
\begin{equation}
\label{eq:sw}
\nu(T) =\nu(0)\left[ 1- a T^{3/2} e^{-E_g/T}\right],
\end{equation}
where $a$ is a fitting parameter.\cite{narath67} The dotted lines in Fig.~\ref{fig:Tspec}
are fits to this expression using $E_g = 17.4$ K.\cite{garlea08} The fit is 
excellent for $T < 15$ K, but deviates above 
$T\gtrsim 0.25 T_N$.  Unfortunately, we
lose the signal near 30 K due to short spin-spin relaxation rates, $1/T_2$,
which increase as thermal fluctuations set in near the phase transition at
$T_N=56$ K.

\begin{figure}
\centering
\includegraphics[width=0.9\linewidth]{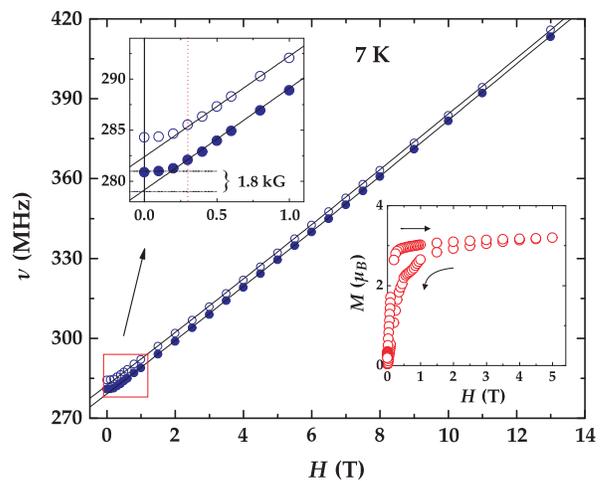}
\caption{(Color online) External field dependence of resonance frequencies of 
two peaks at 7 K. Low 
field data are enlarged in inset.  Zero slope at $H=0$ indicates the existence
of domain structure.  Excellent fit above 0.3 T to 13 T with Eq.~(\ref{eq:hf})
was obtained with the fixed set of parameters (see text), indicating
the formation of single domain above 0.3 T, and the robust spin structure up to
13 T. Inset shows the magnetization $M$ in field along [110] direction at the same 
$T$.  $M$ is saturated  
to 3.2 $\mu_B$ as expected from the non-collinear spin structure. \label{fig:Hspec}
}
\end{figure}

We now turn to the external field dependence of the spectrum.
Figure \ref{fig:Hspec} shows the two peaks as a function of $H$ at 7 K.
The initial slope of $\nu(H)$ vs.~$H$ is zero as seen in the inset of
Fig.~\ref{fig:Hspec}. With increasing $H$, the slope increases gradually and
reaches a fixed value above $H_c \sim 0.3$ T. This behavior indicates the 
existence of the FEM domain structure  
in zero field and its alignment along the external field $H$, forming a single 
domain above $H_c$.  This is also consistent with the 
magnetization $M$ in field that is saturated to 3.2 $\mu_B$ at $H_c$, as shown 
in the inset of Fig.~\ref{fig:Hspec}. Interestingly, $M$ becomes zero as $H\rightarrow 0$.
The absence of the remanent field, despite the ``hysteresis'' in field, 
can be understood by  
assuming the strong coupling  
between the structural and the magnetic domains.  In field, the two domain 
structures can be decoupled because the  
directions of the tetragonal domains are limited to three crystallographic axes 
while the magnetic ones can be aligned in any direction.  Therefore, our data 
suggest that the aligned tetragonal  
domains\cite{suzuki07} along $H$ become random as 
$H\rightarrow 0$, and so do the magnetic ones due to the strong coupling between them.  

The resonance frequency of the V is given by the 
net vector sum of the hyperfine and external fields at the nucleus: 
\begin{equation}
\label{eq:hf}
  \nu(H) = \gamma_N|\mathbf{H}_\text{hf}+\mathbf{H}| = \gamma_N\sqrt{H_\text{hf}^2+H^2+
  2H_\text{hf}H\cos\theta},
\end{equation}
where $\theta$ is the angle between $H_\text{hf}$ and $H$.
This equation is plotted in Fig.~\ref{fig:Hspec} as a solid line, and clearly fits
both peaks up to 13 T with the two parameters $H_\text{hf}$ and $\theta$ for
each peak. We find that the fields and angles are 24.94 T
with $\theta=27.5^\circ$ and 25.23 T with
$\theta=28.8^\circ$.  Since Mn$^{2+}$ is an orbital singlet ion ($3d^5$, $L=0$), 
the Mn$^{2+}$ moment can be easily aligned along $H$ due 
to its large moment\cite{garlea08} (4.2 $\mu_B$) and the essentially 
isotropic susceptibility.  Then the Mn-V and V-V
exchange interactions maintain
the non-collinear FEM spin structure, which is depicted in the inset of Fig.~\ref{fig:spec}.
The excellent fits of data to Eq.~(\ref{eq:hf}) indicate that the non-collinear spin 
structure remains robust up to 13 T. We emphasize that the different 
angles and hyperfine fields at two V sites are inconsistent with either 
\textit{simple} antiferro- or 
ferro-orbital ordering  because the V site are equivalent in both cases. Again, however,
``antiferro'' orbital chains formed by rotating orbitals 
with respect to each other\cite{sarkar09} is compatible with the two 
inequivalent V sites because the hyperfine fields  
at $^{51}$V nuclei can vary depending on the direction of the orbitals and/or 
on different overlaps of the on-site orbital with surrounding V $d$ or O $2p$ 
orbitals. 

We note, however, that the angles we obtain ($\sim 28^\circ$) differ significantly from
those measured by neutron scattering ($\sim 65^\circ$).\cite{plumier89,
garlea08} This result suggests that either (i) the 
orientation of the ordered spins changes in field, or (ii) $\mathbf{H}_\text{hf}$ is not 
coincident with the direction of the ordered V moments.  
Case (i) is possible if the delicate balance of
exchange fields that gives rise to the particular FEM structure in zero field
is modified by the presence of an external field. However, 
case (i) seems unlikely because the magnetization data at 7 K shows the 
saturation moment of about 3.2 $\mu_B$ (inset of Fig.~\ref{fig:Hspec}), which is very close 
to the expected value of 3.1 $\mu_B$ from the given moments and the angle of 
$\sim 65^\circ$.\cite{garlea08}
Therefore, the large tilted 
angles of $\sim 35^\circ$ between $-\mathbf{S}$ and  
$\mathbf{H}_\text{hf}$ requires that the on-site orbital and dipolar terms be 
the same order as the isotropic Fermi contact term that is parallel to 
$-\mathbf{S}$. For the V$^{3+}$ ion, we can estimate the Fermi term $H_F\sim -28$ 
T.\cite{jones65a}
The anisotropic dipolar term can be estimated from the relation  
$H_d=4/7\langle r^{-3}\rangle \mu_B =2/7 \times 125 \langle r^{-3}\rangle_\text{a.u.}$
kG.\cite{kubo69} Using  
$\langle r^{-3}\rangle_\text{a.u.}=3.217$,\cite{abragam70} $H_d\sim 11.4$ T. 
For the orbital term, the magnitude of $H_l$ could be approximated as 
$125 \langle r^{-3}\rangle_\text{a.u.} =40$ T for fully unquenched angular 
momentum.\cite{jones65a} Taking into account the quenching, $H_l$ is expected 
to be the same order as the 
dipolar term, and is not necessary to be parallel to $H_F$, resulting in the total
hyperfine field that is quite off the direction of the ordered moment. 
In this sense, the large 
tilted angle of $\mathbf{H}_\text{hf}$ could be understood as a consequence of the
complex orbital ordering in the orbital triplet ground state of V$^{3+}$ ion, 
since the angular momentum is quenched in the perfect antiferro orbital 
ordering.\cite{suzuki07}


In conclusion, our $^{51}$V NMR study on a single crystal of MnV$_2$O$_4$ reveals the 
two inequivalent V sublattices that imply a complicated orbital ordering 
pattern in the $I4_1/a$ symmetry. Although we cannot 
determine a specific orbital ordering solely by NMR, our data put a strong 
constraint on the possible orbital ordering models. We 
find that the model proposed by Sarkar et   
al.\cite{sarkar09} is quite promising being compatible with our data.  
Also we have shown that the orbital ordering and the non-collinear spin structure are 
robust up to 13 T.  


We thank Hironori Sakai and Stuart E.~Brown for useful discussions and suggestions. 
This work was performed at Los Alamos National
Laboratory under the auspices of the US Department of
Energy Office of Science. Also this work was supported by NSF in-house
research program State of Florida under cooperative agreement DMR-0084173 and 
by the EIEG program at FSU.  

\bibliography{/mydocuments/mydata/mybib}

\end{document}